\documentclass[english,aps,prper,reprint,showpacs,titlepage,longbibliography]{revtex4-2}

\usepackage[T1]{fontenc}	
\usepackage[latin9]{inputenc}	
\usepackage{geometry}		
\usepackage{graphicx}
\usepackage[above,below]{placeins}	
\usepackage{times}
\usepackage{amsmath}
\usepackage{array}

\usepackage{float}

\usepackage{amsfonts} 
\usepackage{siunitx} 
\usepackage{color}
\usepackage{mathrsfs}
\usepackage{multirow}

\usepackage{verbatim}
\usepackage{hyperref}  
\hypersetup{colorlinks=true,urlcolor=blue,citecolor=blue,linkcolor=blue}   
\usepackage{enumitem}          
\setlist{nosep}                 

\begin{document}

\begin{titlepage}

\title{Real-time quantitative measurement of a Stirling engine $P$-$V$ diagram}

\author{Nikolai I.\ Lesack}
\author{Hiroko Nakahara}
\author{Jake S.\ Bobowski}	
\affiliation{Department of Physics, University of British Columbia, 3333 University Way, Kelowna, British Columbia, V1V 1V7,
    Canada} 


\begin{abstract}
This paper describes simple modifications to a Stirling-type heat engine that allow its $P$-$V$ diagram to be measured.  The main advantage of our approach is that a calibrated $P$-$V$ diagram can be measured and displayed in real time as the engine is running.  Our implementation uses a relatively inexpensive, but high-quality, gamma-type Stirling engine designed for demonstrations.  The only modifications required to the as-purchased engine are a single hole drilled in the top plate to accommodate a pressure sensor and attaching circular choppers to the flywheel.  An outer chopper is used to reset the detection electronics when the internal volume of the engine is a minimum. An inner chopper is then used to track the orientation of the flywheel.  This paper describes the design of the photogates and electronics used to collect and process the data.  Example data are shown to highlight the capabilities of the system.
\end{abstract}  

 \maketitle
\end{titlepage}
\section{Introduction}
At the University of British Columbia, we have instrumented demonstration Stirling engines in order to make quantitative measurements of their $P$-$V$ diagrams.  The gas pressure is monitored by drilling a single hole through the top plate of the engine to accommodate an inexpensive differential pressure sensor.  In the first design, the volume was measured using a reflector and an IR emitter/receiver (QRD1114).  The reflector was attached to the engine's piston and used to redirect the emitted IR signal back towards the receiver which outputs a voltage that varies nonlinearly with the reflector-receiver separation distance.  The advantage of this design was that a calibrated $P$-$V$ diagram could be recorded and displayed  while the engine was running~\cite{Nakahara:2012}.  Some of the disadvantages include the added weight to the piston, some additional machining of the top plate, the nonlinear calibration factor, and the need to use software to read the output of the IR sensor and calculate the corresponding gas volume.  

In 2020, we introduced a second design in which the height of the piston was monitored using a photogate and chopper that was printed onto an acetate sheet and then taped onto the engine's flywheel.  This design eliminated the need for a reflector and additional machining~\cite{Lu:2020}.  However, in that system, it was necessary to first collect the data and then process it afterwards to construct the $P$-$V$ diagram.  
In this paper, we describe a variation of the chopper design that, when combined with some relatively simple electronics, can be used to generate calibrated $P$-$V$ diagram data in real time.

\begin{figure}
    \centering{
        \begin{tabular}{rlcrl}
            (a) & \includegraphics[keepaspectratio, height=0.55\linewidth]{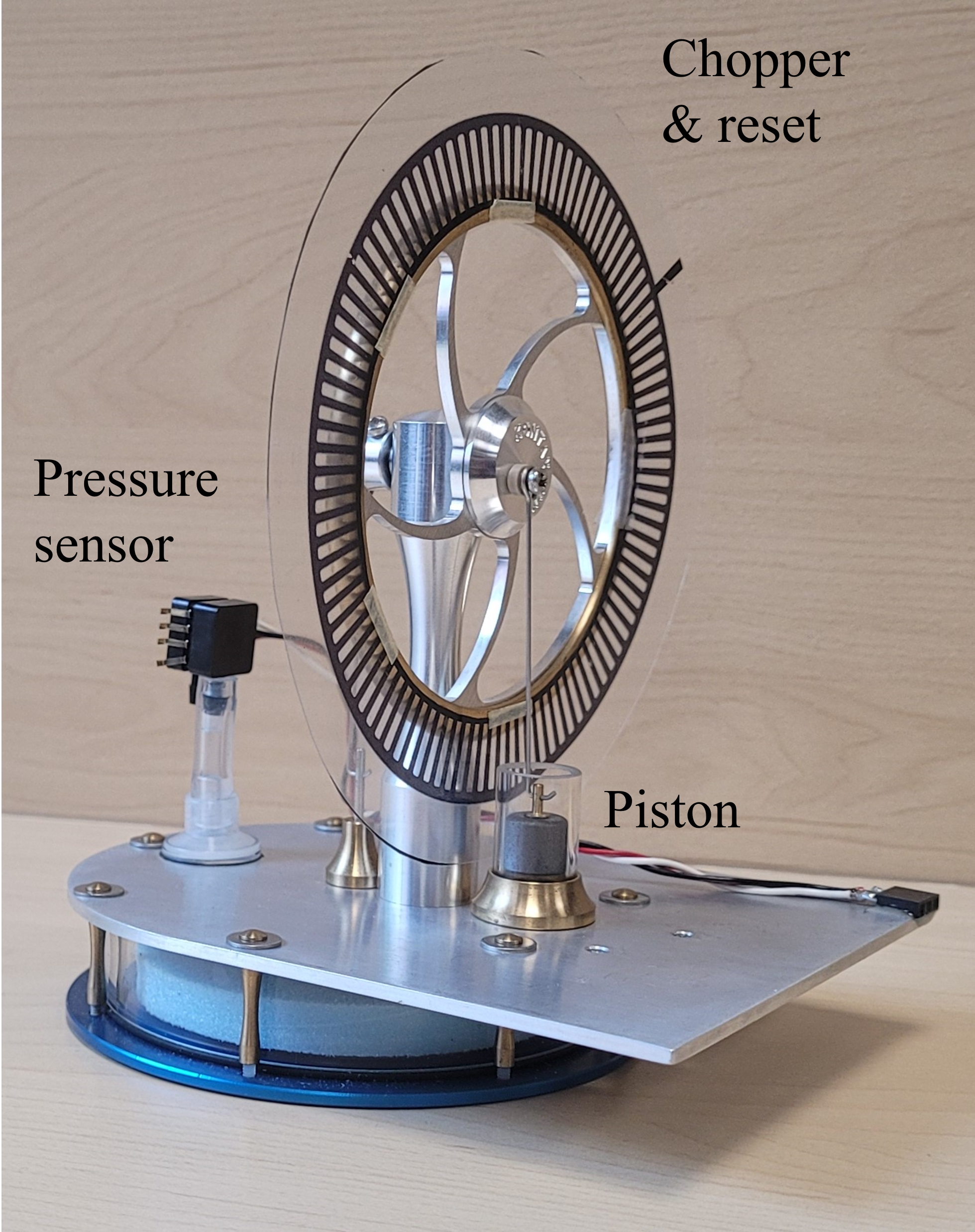} & \quad & (b) \includegraphics[keepaspectratio, height=0.55\linewidth]{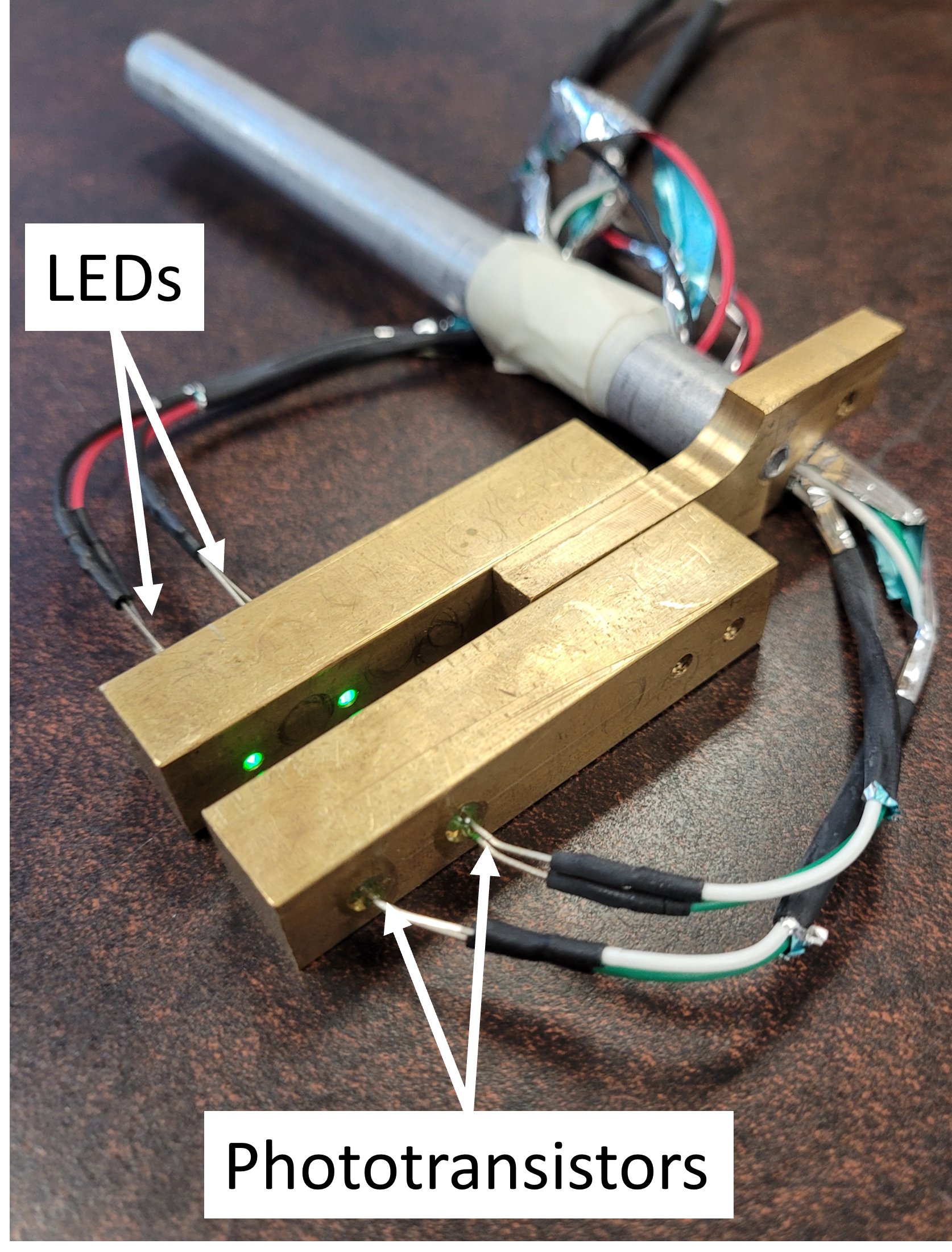}\\
            ~ & ~\\
            (c) & \multicolumn{4}{c}{\includegraphics[keepaspectratio, width=0.9\linewidth]{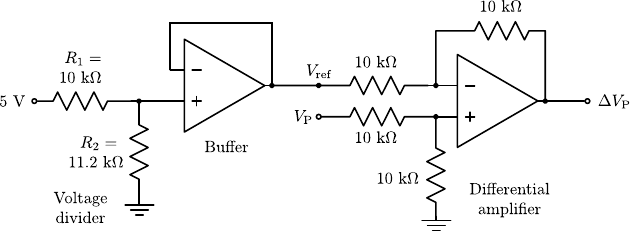}}
        \end{tabular}
    }
    \caption{\label{fig:Stirling}(a) Photograph of the instrumented Stirling engine showing the choppers attached to the flywheel and the differential pressure sensor.  (b) Photograph of the home-built dual photogate.  (c) The op-amp circuit used to calculate the change in pressure.}
\end{figure}

\begin{figure*}
\centering{
	\begin{tabular}{cc}
		(a) & \includegraphics[keepaspectratio, width=0.8\linewidth]{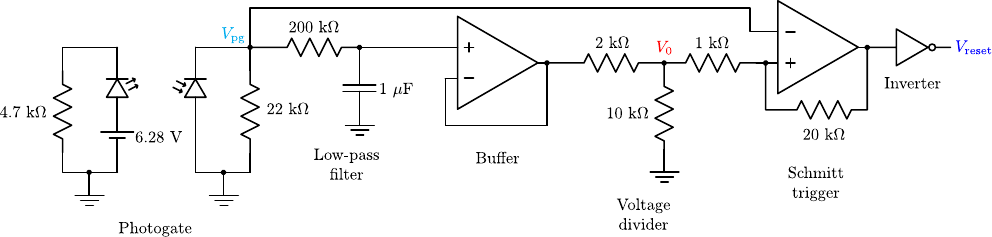}\\
        ~ & ~\\
		(b) & \includegraphics[keepaspectratio, width=0.75\linewidth]{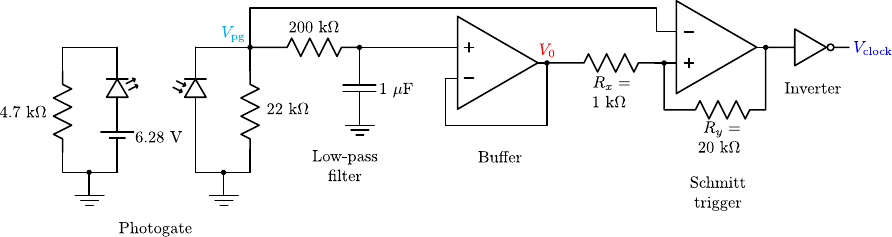}\\
        ~ & ~\\
        (c) & \includegraphics[keepaspectratio, width=0.9\linewidth]{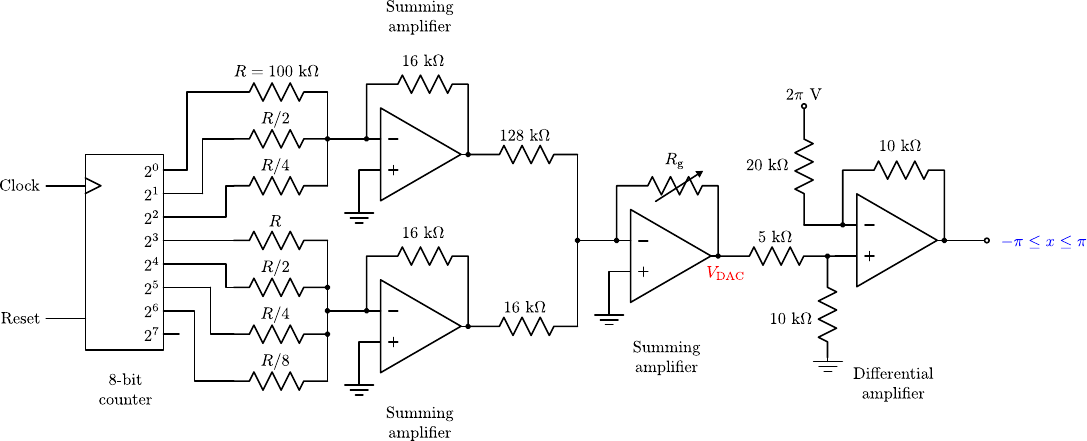}
	\end{tabular}
}
\caption{\label{fig:circuits}(a) The circuit used to send ``reset'' pulses to the binary counter.
(b) The circuit used to send ``clock'' pulses to the binary counter. (c) The digital-to-analog converter used to determine the orientation $x$ of the flywheel.   
}
\end{figure*}

\section{Design \& Electronics}\label{sec:electronics}
The Stirling engine used in our experiments was the KS90 manufactured by Kontax Stirling Engines~\cite{Kontax}. Figure~\ref{fig:Stirling}(a) shows a photograph of the instrumented Stirling engine.  The pressure sensor used is the MPXV7002 from NXP Semiconductors.  It outputs a voltage that varies linearly between $0.5$ to $4.5\rm\ V$ as the pressure changes from $-2$ to $2\rm\ kPa$~\cite{MPXV7002}.  Figure~\ref{fig:Stirling}(c) shows the circuit used to calculate the change in pressure of the system.  Our particular sensor output a voltage of $2.65\rm\ V$ when there was no pressure difference across the inlet and outlet.  We first used a voltage divider and buffer to construct \mbox{$V_\mathrm{ref} = 5~\mathrm{V}\,\left[R_2/(R_1 + R_2)\right]\approx 2.65\rm\ V$} and then used a differential amplifier to find \mbox{$\Delta V_\mathrm{P} = V_\mathrm{P} - V_\mathrm{ref}$}, where $V_\mathrm{P}$ is pressure sensor output~\cite{Snoke:2015}.  The required $5\rm\ V$ DC was provided by an AS431 voltage regulator.  This same $5\rm\ V$ was also used to power the pressure sensor.  All of the operational amplifiers (op-amps) used in the circuits presented in this paper were TL031 JFET-input integrated circuits (ICs).  Because the pressure sensor sensitivity is $1\rm\ V/kPa$, $\Delta V_\mathrm{P}$ can be interpreted as the change in pressure in units of kilopascals.  

Figure~\ref{fig:Stirling}(a) also shows a pair of choppers that have printed onto an acetate sheet and then taped to the flywheel of the Stirling engine.  The outer chopper has only a single narrow opaque section while the inner chopper is divided into $100$ equally-spaced transparent and opaque intervals.  The outer chopper is used to reset or clear a counter IC when the piston is at the bottom of its stroke. 
The inner chopper is then used to increment the count from zero to $99$ with each rotation of the flywheel.  Finally, a digital-to-analog converter (DAC) converts the binary output of the counter to an analog voltage that varies between $x = -\pi$ and $+\pi\rm\ V$, with $x=\pm\pi$ corresponding to the orientation at which the volume is a minimum and $x=0$ corresponding to the maximum volume.   

\begin{figure}[ht!]
\centering{
	\begin{tabular}{rl}
		(a) & \includegraphics[keepaspectratio, width=0.72\columnwidth]{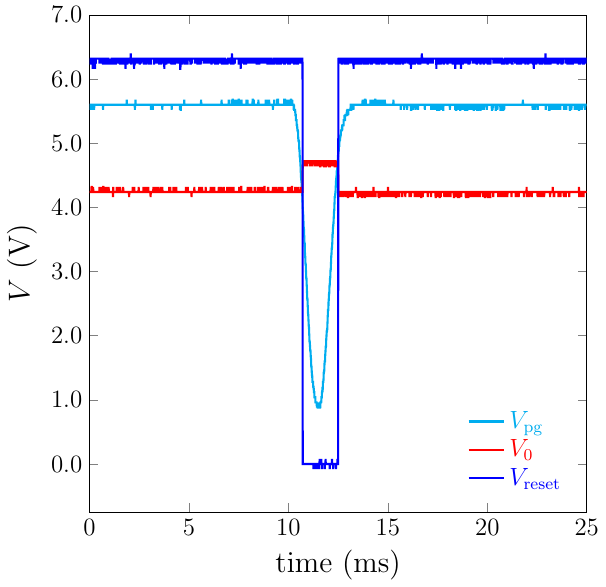}\\
        ~ & ~\\
		(b) & \includegraphics[keepaspectratio, width=0.72\columnwidth]{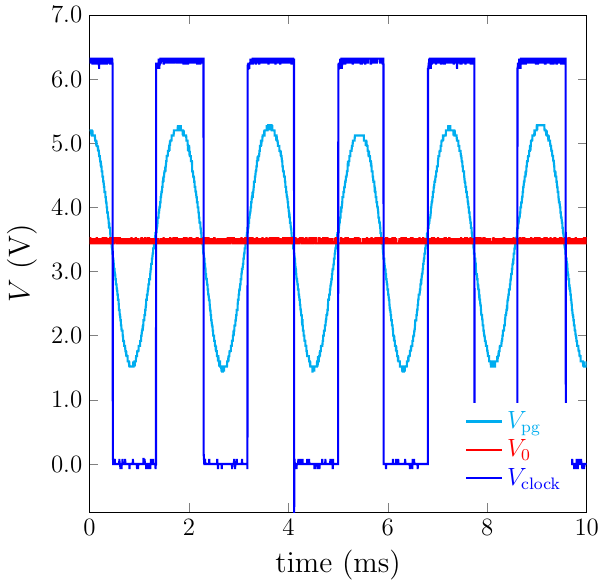}\\
        ~ & ~\\
        (c) & \includegraphics[keepaspectratio, width=0.72\columnwidth]{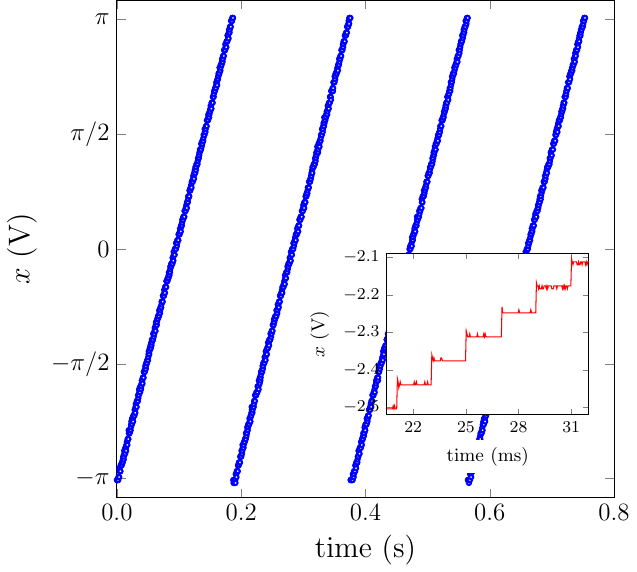}
	\end{tabular}
}
\caption{\label{fig:circuitData}(a) Data collected from the reset circuit of Fig.~\ref{fig:circuits}(a).
(b) Data collected from the chopper circuit of Fig.~\ref{fig:circuits}(b).  (c) Data from the DAC circuit of Fig.~\ref{fig:circuits}(c).  The inset shows a detailed view of $x$ versus $t$ highlighting the discrete steps in its time evolution.   
}
\end{figure}

Figure~\ref{fig:Stirling}(b) shows a dual-photogate designed in-house and used to interrogate the states of the choppers. As Figs.~\ref{fig:circuits}(a) and (b) show, the LEDs were forward biased using a voltage of $6.28\rm\ V$ (provided by an AS431 voltage regulator) and a $4.7\rm\ k\Omega$ resistor.  The receiver was made from an NPN phototransistor and a $22\rm\ k\Omega$ resistor.  The $22\rm\ k\Omega$ resistor ensured that the change in voltage between the blocked and unblocked states was large while still maintaining a sufficiently-high response time.  The receiver output is taken to be the voltage across the $22\rm\ k\Omega$ resistor.  When the light incident on the phototransistor is blocked, it is not conducting and the output voltage is low.  On the other hand, when the light from the LED is not blocked, the phototransistor conducts current and a larger voltage develops across the $22\rm\ k\Omega$ resistor.  Figs.~\ref{fig:circuitData}(a) and (b) show the measured photogate voltages $V_\mathrm{pg}$ while using a cup of boiled water to run the Stirling engine.

As shown in Figs.~\ref{fig:circuits}(a) and (b), the photogate signals were conditioned using a low-pass filter and Schmitt trigger.  The frequency of the $V_\mathrm{pg}$ signal from the ``clock'' pulse circuit is approximately $550\rm\ Hz$.  The Gaussian-like reset pulse has a width of $\sigma_t\approx 0.5\rm\ ms$ in the time domain which corresponds to a width of $\sigma_f=(2\pi\sigma_t)^{-1}\approx 320\rm\ Hz$ in the frequency domain. 
Passing these signals through a low-pass filter with a corner frequency of $(2\pi RC)^{-1}\approx 0.80\rm\ Hz$ effectively averages the signals over many cycles and returns the DC offset.  In the clock pulse circuit, this offset is denoted $V_0$ and is used as a Schmitt trigger threshold voltage. Figure~\ref{fig:circuitData}(b) shows that, as expected, $V_0$ is located at the average value of $V_\mathrm{pg}$.  

In the case of the reset circuit, the average value of $V_\mathrm{pg}$ is very close to the unblocked voltage of $V_\mathrm{pg}$ and would not be suitable as a Schmitt trigger threshold voltage.  For this reason, a voltage divider is used to reduce the output of the low-pass filter before $V_0$ is passed to the Schmitt trigger.  As Fig.~\ref{fig:circuitData}(a) shows, the value of $V_0$ changes abruptly when the Schmitt trigger changes state.  Although this issue does not affect the performance of the circuit, it could be avoid by inserting a buffer between the voltage divider and the $1\rm\ k\Omega$ resistor leading to the Schmitt trigger's non-inverting input.

The op-amps used as Schmitt triggers were powered with $+2\pi\rm\ V$ and $-1\rm\ V$ such that their high and low state outputs were $V_\mathrm{H}\approx 5.28\rm\ V$ and $V_\mathrm{L}\approx 0\rm\ V$, respectively.  Therefore, for the circuits of Figs.~\ref{fig:circuits}(a) and (b), the low-to-high transition is expected to occur when
\begin{equation}
    V_\mathrm{pg}<V_\mathrm{L\to H} = V_0\frac{R_y}{R_x + R_y} + V_\mathrm{L}\frac{R_x}{R_x + R_y},
\end{equation}
while the high-to-low transition occurs when
\begin{equation}
    V_\mathrm{pg}>V_\mathrm{H\to L} = V_0\frac{R_y}{R_x + R_y} + V_\mathrm{H}\frac{R_x}{R_x + R_y}.
\end{equation}
Since $V_\mathrm{L}\approx 0$, the difference between the two transition threshold voltages is \mbox{$V_\mathrm{H}R_x/(R_x + R_y)\approx 0.25\rm\ V$}.  Finally, the Schmitt trigger outputs are passed through inverters before being sent to the reset and clock inputs of an 8-bit counter.  We used an SN54HC590 counter which clears the outputs when the reset signal is low and advances the count on the low-to-high transitions of the clock signal.

The counter counts from zero to $99$ with each rotation of the flywheel.  A DAC is used to convert the binary output of the counter to an analog voltage.  As shown in Fig.~\ref{fig:circuits}(c), we designed and built our own DAC using inverting summing amplifiers, although it is also possible to purchase an inexpensive 8-bit DAC.  Based on the component values given in the figure, the output of our DAC is given by
\begin{multline}    
    V_\mathrm{DAC} = \frac{R_\mathrm{g}}{8}\left(\frac{V_{2^0}}{100} + \frac{V_{2^1}}{50} + \frac{V_{2^2}}{25}\right) \\
    + R_\mathrm{g}\left(\frac{V_{2^3}}{100} + \frac{V_{2^4}}{50} + \frac{V_{2^5}}{25} + \frac{V_{2^6}}{12.5}\right),
\end{multline}
where $V_{2^n}$ represents the digital outputs of the counter~\cite{Snoke:2015, Horowitz:2015}.  The value of $R_\mathrm{g}$ was chosen such that, when the counter reaches a terminal count of $99$ ($1100011$ in binary), \mbox{$V_\mathrm{DAC}=2\pi(99/100)\rm\ V$}.

Finally, for our application, we wanted a voltage $x$ that varied from $-\pi$ to $+\pi$ rather than from zero to $2\pi$.  The differential amplifier in Fig.~\ref{fig:circuits}(c) was used to subtract $\pi\rm\ V$ from $V_\mathrm{DAC}$ to produce a final output $-\pi<x<\pi$ that tracks the orientation of the of the Stirling engine flywheel.  Figure~\ref{fig:circuitData}(c) shows $x$ as a function of time for four rotations of the flywheel.  The inset shows a zoomed-in view of $x$ versus time, highlighting the discrete changes in its value each time a clock pulse increments the output of the counter.

\section{Calculating the change in volume}\label{sec:V}

\begin{figure}[t]
\centering{
\includegraphics[keepaspectratio, width=0.83\columnwidth]{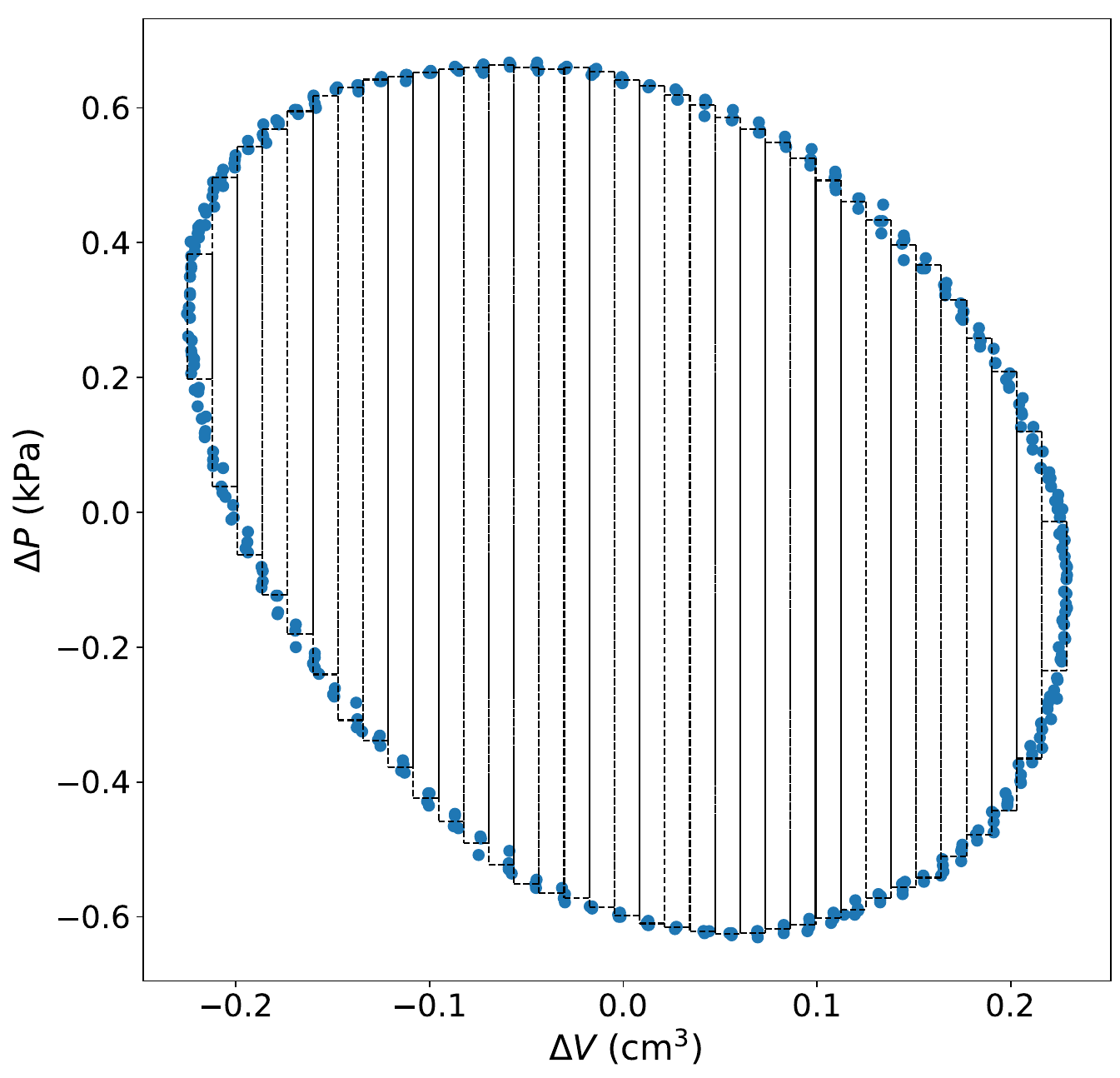}
}
\caption{\label{fig:area}Output of the Python code used to estimate the area enclosed by a measured $P$-$V$ diagram.   
}
\end{figure}

As described in Ref.~\cite{Lu:2020}, the change in internal volume of the Stirling can be expressed as
\begin{equation}
    \Delta V \approx Ar\left(\cos x + \dfrac{r}{2\ell}\sin^2 x\right),\label{eq:dV}
\end{equation}
where $A$ is the cross-sectional area of the piston, $\ell$ is the length of the rod that connects to the piston to the flywheel, and $r$ is the distance between the flywheel's rotation axis and the connecting rod's anchor point.  See Fig.~1(b) in Ref.~\cite{Bobowski:2023}.  One option is to use a data acquisition device (DAQ) to read in the value $x$ and then complete the calculation of $\Delta V$ in software.  There are numerous options that will allow the data acquisition and calculations to be done in real time.  Some that we're familiar with include a National Instruments DAQ paired with LabVIEW, Vernier voltage sensors paired with LoggerPro, and the Arduino platform. 

Alternatively, as described in Ref.~\cite{Bobowski:2023}, it is possible to design  additional circuits to evaluate $\cos x$ and $\sin x$.  An analog multiplier and summing amplifier can then be used to find $\Delta V$ as given by Eq.~(\ref{eq:dV}).  The advantage of this method is that, using the $XY$-mode of an oscilloscope, calibrated $P$-$V$ diagrams can be displayed in real time without relying on other DAQ devices or software packages.  

Figure~\ref{fig:area} shows an example set of $\Delta P$ versus $\Delta V$ data collected using the Stirling engine shown in Fig.~\ref{fig:Stirling}(a).  The area enclosed by the $P$-$V$ diagram represents the work done per cycle by the engine.  We have written a Python script to approximately calculate this area~\cite{Lesack:2022}.  The script starts by dividing the area into $N$ bins of equal width.  Next, a rectangular area is assigned to each bin by first averaging the $\Delta P$ values of the points that fall within a particular bin to find $\mu_i$.  The all points above the value of $\mu_i$ are then averaged to define the position of the top of the rectangle.  The bottom of the rectangle is found by averaging the $\Delta P$ values of the points that fall below $\mu_i$.  Finally, with the height and width of each rectangle known, the area of the $P$-$V$ diagram is approximated by summing the areas of all the rectangles.  Figure~\ref{fig:area} shows the rectangles used to fill the area of the measured $P$-$V$ diagram.  Using the areas of these rectangles, the work done per cycle of the Stirling engine was found to be $435\rm\ \mu J$.

\section{Summary}
Several authors have recently written about innovative undergraduate lab activities that make use of inexpensive demonstration Stirling engines~\cite{Lu:2020, Yeadon:2021, Stevens:2023}.  In this work, we have described simple modifications to an as-purchased Stirling engine that allow accurate $P$-$V$ diagrams to be recorded and displayed in real time as the engine is running.  Changes in the gas pressure are monitored using a differential pressure sensor and the position of the piston that sets the gas volume is monitored using photogates and choppers that have been attached to the engine's flywheel.  This paper has focused on the electronics that we designed to process the data collected from the pressure sensor and photogates.

\end{document}